\documentclass[aps,rmp,reprint,superscriptaddress]{revtex4-2}
\usepackage{yfonts, amssymb, graphicx, amsfonts,epstopdf }

\usepackage[dvipsnames]{xcolor}
\usepackage{multirow,amssymb,amsbsy,amsmath,epstopdf}
\usepackage[colorlinks,allcolors=blue]{hyperref}
\setcitestyle{super,sort&compress,comma}
\usepackage{graphicx}
\usepackage{verbatim}
\usepackage{bm}
\usepackage{bbold}
\usepackage{amsmath}

\begin{document}

\title{Counterfactual communication without a trace in the transmission channel}

\author{Wei-Wei Pan}
\affiliation{CAS Key Laboratory of Quantum Information, University of Science and Technology of China, Hefei 230026, China}
\affiliation{CAS Center for Excellence in Quantum Information and Quantum Physics, University of Science and Technology of China, Hefei 230026, China}
\affiliation{School of Physics and Materials Engineering, Hefei Normal University, Hefei, Anhui 230601, China.}

\author{Xiao Liu}
\author{Xiao-Ye Xu}
\email{xuxiaoye@ustc.edu.cn}
\author{Qin-Qin Wang}
\affiliation{CAS Key Laboratory of Quantum Information, University of Science and Technology of China, Hefei 230026, China}
\affiliation{CAS Center for Excellence in Quantum Information and Quantum Physics, University of Science and Technology of China, Hefei 230026, China}
\affiliation{Hefei National Laboratory, University of Science and Technology of China, Hefei 230088, China}

\author{Ze-Di Cheng}
\affiliation{CAS Key Laboratory of Quantum Information, University of Science and Technology of China, Hefei 230026, China}
\affiliation{CAS Center for Excellence in Quantum Information and Quantum Physics, University of Science and Technology of China, Hefei 230026, China}

\author{Jian Wang}
\author{Zhao-Di Liu}
\author{Geng Chen}
\author{Zong-Quan Zhou}
\affiliation{CAS Key Laboratory of Quantum Information, University of Science and Technology of China, Hefei 230026, China}
\affiliation{CAS Center for Excellence in Quantum Information and Quantum Physics, University of Science and Technology of China, Hefei 230026, China}
\affiliation{Hefei National Laboratory, University of Science and Technology of China, Hefei 230088, China}

\author{Chuan-Feng Li}
\email{cfli@ustc.edu.cn}
\author{Guang-Can Guo}
\affiliation{CAS Key Laboratory of Quantum Information, University of Science and Technology of China, Hefei 230026, China}
\affiliation{CAS Center for Excellence in Quantum Information and Quantum Physics, University of Science and Technology of China, Hefei 230026, China}
\affiliation{Hefei National Laboratory, University of Science and Technology of China, Hefei 230088, China}

\author{Justin Dressel}
\affiliation{Institute for Quantum Studies, Chapman University, Orange, CA 92866}
\affiliation{Schmid College of Science and Technology, Chapman University, Orange, CA 92866}

\author{Lev Vaidman}
\email{vaidman@tauex.tau.ac.il}
\affiliation{Institute for Quantum Studies, Chapman University, Orange, CA 92866}
\affiliation{Raymond and Beverly Sackler School of Physics and Astronomy, Tel-Aviv University, Tel-Aviv 69978, Israel}
\affiliation{Max-Planck-Institut f\"{u}r Quantenoptik, Hans-Kopfermann-Stra{\ss}e 1, 85748 Garching, Germany}

\date{\today}

\begin{abstract}
We report an experimental realization of a modified counterfactual communication protocol that eliminates the dominant environmental trace left by photons passing through the transmission channel. Compared to Wheeler's criterion for inferring past particle paths, as used in prior protocols, our trace criterion provide stronger support for the claim of the counterfactuality of the communication.
We verify the lack of trace left by transmitted photons via tagging the propagation arms of an interferometric device by distinct frequency-shifts and finding that the collected photons have no frequency shift which corresponds to the transmission channel. As a proof of principle, we counterfactually transfer a quick response code image with sufficient fidelity to be scanned with a cell phone.
\end{abstract}

\maketitle
\newpage

\yinipar{\textcolor{BurntOrange}{A}}ccording to classical physics, information must be transferred from one site to another by a local carrier along a continuous path. The carrier might be a messenger, an electric pulse in a wire, or the photons of a cellular phone communication channel. Transfer of information without particles being detectable in the transmission channel would defy common sense. Nevertheless, quantum counterfactual communication \cite{Kwiat,SalihCFC} claims to do exactly this, so unsurprisingly, it has been a controversial topic. Proponents have gone so far as to claim even quantum information, such as a quantum state, may be transferred counterfactually \cite{SalihQCF,ZubairyQCF,Salih2022}.
	
These counterfactual communication protocols crucially rely upon interaction-free detection (IFD), which was originally introduced in a paper \cite{IFM} co-authored by Vaidman, who is also an author of the present paper. However, Vaidman has repeatedly noted that IFD can only verify the existence of an object at a location, not its absence.
 This corresponds to a one-valued counterfactual transfer which is  enough for counterfactual key distribution \cite{Noh,LiuKD}, but does not allow general counterfactual computation \cite{Kwiat}, communication of classical data \cite{SalihCFC}, or quantum state transfer \cite{SalihQCF,ZubairyQCF}. For these proposals and their experimental implementation \cite{expPNAS}, Vaidman showed \cite{V07,V14,V15} that 
 when the absence of an object in a particular place is established, the particle leaves a trace en route similar to the trace of a single localized particle.

\begin{figure}[t]
    \centering
    \includegraphics[width = 0.85\columnwidth]{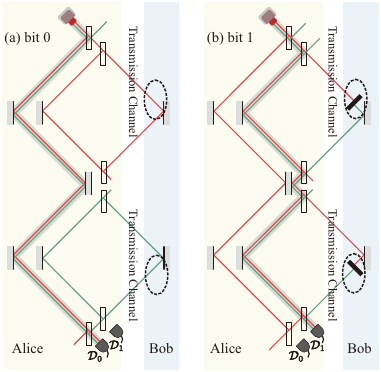}
    \caption{\textbf{Counterfactual communication, unfolded.} Alice send single photons into  nested MZIs connected by a double-sided mirror. Bob sends bit 1 to Alice by placing two shutters, one in $A_1$ and another in $A_2$. Bob sends  bit 0 by leaving the interferometer undisturbed. A trace will only be observed where the forward- (red) and backward-evolving (green) paths overlap (thicker gray line), and is absent from the transmission channel for both bit values. (a) Bob sets a value of 0 by leaving the inner MZIs unblocked. The inner MZIs are tuned to destructively interfere and cancel the path to the connecting mirror, so collected photons can only trigger detector $D_0$ for Alice. (b) Bob sets a value of 1 by inserting two shutters. The shutters  change the interference so that the photon can only trigger $D_1$.
    }
    \label{fig.sketcha}
\end{figure}

It is simple to show why it impossible to devise a protocol that unambiguously distinguishes between blocking and not blocking one particular location $A$ without leaving a first-order trace in that location. Alice sends a photon that can be detected in detector $D_0$ or $D_1$. Placing an object in the location $A$ must make detection in $D_0$ impossible. By our rules, the object can only absorb and not reflect, so this change tells us that there is a path through $A$ from Alice's source to the detector $D_0$. Therefore, there have to be a path from the source to $A$ and also a path from  $D_0$ to $A$. In such a situation, the two-state vector formalism (TSVF) \cite{TSVF}, which adds to the description of the photon a quantum state evolving backward in time from the postselection at the detector, provides a simple way to find the location of the weak traces left by the photon.
The overlap of the forward and backward evolving states of the photon at $A$ leads to the trace in $A$ of the first order in the interaction parameter \cite{past,PNAS}, similar to the trace of a photon localized at $A$. This makes it unreasonable to claim that the photon was not there, i.e. that this was a counterfactual communication.

The experiment in this paper addresses this critical weakness of earlier protocols. We implement a protocol design based on the modification of these protocols found recently by Aharonov and Vaidman (AV)  \cite{AV19} that eliminates the trace of the first order also in the case of absence of the object, therefore permitting communication that is counterfactual according to our more stringent weak trace criterion. 

The key ingredient of the AV modification of counterfactual protocols is that Bob communicates his bit to Alice by blocking or not blocking {\it two} locations $A_1$ and $A_2$ together, instead of one location $A$, see Fig. 1. 
Alice gets a click in detector $D_0$ when $A_1$ and $A_2$ are not blocked, but it must be impossible when locations $A_1$ and $A_2$  are blocked.  Therefore, there must be paths from the source to the detector through $A_1$ and $A_2$ that cause destructive interference of the waves from all the paths together. There are three such paths: through $A_1$ and $A_2$,   through $A_1 $ but not $A_2$, and through $A_2 $ but not $A_1$. 
These additional paths provide a finite contribution of the amplitude of the photon wave in the detector $D_0$. However, the waves in these three paths interfere destructively in both $A_1$ and $A_2$. 
There is no overlap of the forward and backward evolving wave functions at $A_1$ and $A_2$, and consequently, no first order trace is created. Note that blocking only one of the locations $A_1$ or $A_2$, does not lead to a possibility of detection at $D_0$, so the argument of non-counterfactuality of the non-modified protocols does not hold here.

In our experiment, we do not only perform the counterfactual communication protocol with AV modification, but we also  verify that both logical bit values are communicated without leaving a significant trace in the connecting path of the transmission channel. By applying small and distinct frequency shifts to each photon at different locations and filtering by frequency during collection \cite{AskingphotonKedem,Danan}, we show where those photons interacted significantly to leave a trace and where they did not. We verify that any trace left by information-carrying photons in the transmission channel is small enough to not be visible above the noise floor, unlike the much larger traces left by single photons during calibration.

Our experimental results are a proof of principle. The transferred bits have high enough fidelity to send a QR code image that can be scanned by a cell phone, but the current design has a small success probability for the sent photons to be detected.
  Since the stray photons may be intercepted, this simple design is not yet secure; however, the  application of our modification to counterfactual protocols based on the quantum Zeno effect \cite{expPNAS} could dramatically increase the success probability,  potentially enabling secure communication.

\begin{figure}[t]
    \centering
    \includegraphics[width = 1\columnwidth]{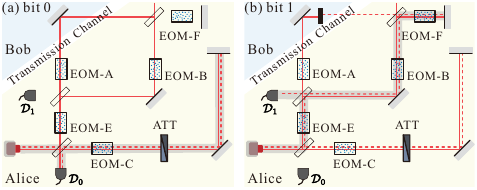}
    \caption{\textbf{Folded design as MMI with nested MZI.} Only the forward-evolving state is shown, with the solid red line becoming dashed after bouncing off the mirror at the end of the lower path.
        (a) With no blocks, only $D_0$ can click to record a bit value of 0. The photons entering the MZI leave the interferometer, so $D_1$ cannot click.
        (b) When Bob is blocking, only $D_1$ can click to record a bit value of 1. Destructive interference prevents clicks at $D_0$. Neither case has an overlap path (gray line) in the transmission channel, so no trace of EOM-A will be seen in the frequencies of collected photons.}
    \label{fig.sketchb}
\end{figure}

\begin{figure*}
    \centering
    \includegraphics[width = 0.95\textwidth]{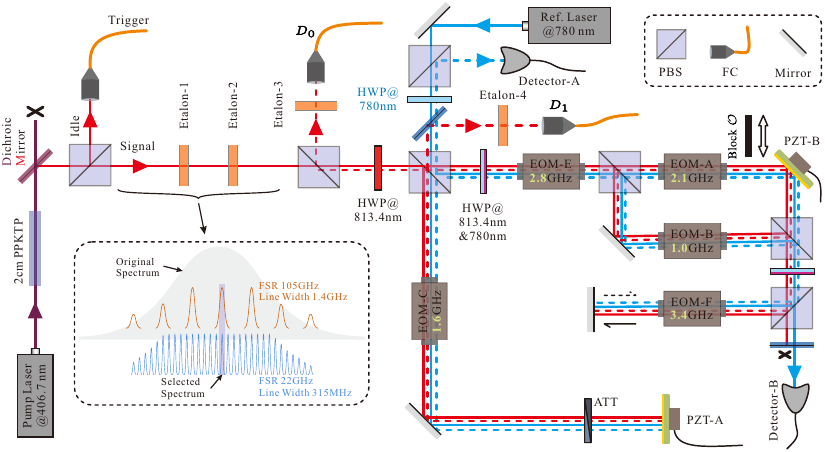}
\caption{\textbf{Experimental Setup.}
A 1\,mW continuous wave (CW) laser centered at 406.7\,nm pumps a 2\,cm long PPKTP crystal, generating degenerate photon pairs. The residual pump laser is removed by a dichroic mirror\,(DM). The idle and signal photons both have wave lengths centered at 813.4\,nm but have opposite polarizations. After separation by a polarization beam splitter\,(PBS), the idle photons enter a fiber collimator (FC) and trigger a single photon detector that heralds corresponding time-tagged signal photons entering the folded double-nested MZI. 
A reference laser (CW at 780\,nm) independently traverses the interferometers to actively stabilize their phases; i.e., the electrical signals from the Detectors A and B are fed back to control corresponding piezoceramic\,(PZT)-driven mirrors and implement phase-locking. We spectrally filter the reference photons from the pump photons before detection to maintain isolation.
To implement unbiased beam splitters and remove residual polarization effects, we insert half-wave plates\,(HWPs) oriented with optical axes at $22.5^\circ$ before each PBS. We also insert an attenuator\,(ATT) as needed to balance the optical intensities of the MMI arms and maximize interference visibility. 
Tracing the paths taken by the photons then proceeds in three steps: First, we use two etalons---Etalon-1 with free spectral range (FSR) of 105\,GHz and line width of 1.4\,GHz, and Etalon-2 with FSR 22\,GHz and line width 315\,MHz---to prepare narrow-band single photons as shown schematically in the dashed box.
Second, we insert five electro-optic modulators\,(EOMs), labeled EOM-A, -B, -C, -E, and -F, with distinguishable modulation frequencies 2.1\,GHz, 1.0\,GHz, 1.6\,GHz, 2.8\,GHz, and 3.4\,GHz, respectively, into distinct paths to record where photons have gone by small shifts in their spectral profiles.
Third, we analyze the spectra of the collected photons using the tunable Etalon-3 and Etalon-4 (both with FSR 8 GHz and linewidth of 100 MHz) placed before detectors $D_0$ and $D_1$, such that spectrally-resolved intensities reveal which paths the photons must have taken.}
\label{fig:setup}
\end{figure*}

To explain the basic mechanism for achieving counterfactual communication, we first consider the setup \cite{AV19} shown in Fig.~\ref{fig.sketcha}. It has a nested pair of Mach-Zehnder interferometers (MZIs) that will be conceptually folded onto each other to obtain the compact Michelson-Morely interferometer (MMI) shown in Fig.~\ref{fig.sketchb}, which our experiment implements with heralded single photons and active stabilization as shown in Fig.~\ref{fig:setup}.

The left path in Fig.~\ref{fig.sketcha} is one arm of an (outer) MZI, so each photon emitted by the source at the top left bounces off three mirrors before reaching the final beamsplitter at the bottom, with detector $D_0$ placed at the right output port. We then split the right arm of this outer MZI into two consecutive (inner) MZIs, with detector $D_1$ placed at the right output port of the second inner MZI. As shown in  Fig.~1(a), we tune each inner MZI to destructively interfere at the double-sided mirror that connects them, allowing all photons in the right arm to escape before they can trigger detector $D_1$. Thus, each photon must either escape or cause the detector $D_0$ to click. However, if we block the right arms of the inner MZIs with shutters, as shown in Fig.~1(b), the remaining unblocked path will either trigger detector $D_1$ or complete the outer MZI, which we tune to destructively interfere at its right output and thus prevent the detector $D_0$ from clicking.

The unfolded design in Fig.~\ref{fig.sketcha} also illustrates both forward- (red) and backward-evolving (green) wave function paths, such that their overlap predicts which traces may be seen in the experiment. In regions where the two overlap (gray), a weakly interacting probe will record a trace consistent with those observed for definite particle paths. However, the same probe will not register an appreciable trace where there is no overlap. Notably, this simple picture predicts that no trace will be observed in the right arms of the inner MZIs, even though they form the transmission channel for information about the block.

For simplicity of implementation, we conceptually fold the design in Fig.~\ref{fig.sketcha} over a horizontal line on the double-sided mirror to identify both inner MZI and obtain Fig.~2. This fold converts the outer MZI into an MMI so that each photon passes through the same inner MZI twice. The benefit of this compact variation is that only one shutter is needed to change the bit configuration, since each photon may reach that shutter from two different directions. The penalty is that the transmissivity of the beamsplitters must also be slightly adjusted when the shutter is inserted to preserve the visibility of the destructive interference. For simplicity, we use 50/50 beamsplitters and balance the intensities with a tunable attenuator placed in the lower arm ($C$).

To interrogate each photon about where it has been without spoiling the sensitive interference, we arrange for the photon to acquire small and distinct frequency-shifts when traversing different arms of the interferometer. We then spectrally filter those photons before detection to obtain frequency-dependent intensities that reveal their past interaction history. Following Zhou et al.\cite{AskingphotonKedem} we use free-space frequency electro-optic modulators (EOM) to induce frequency-shifted sidebands to mark the paths.

We provide additional technical detail about our implementation of Fig.~\ref{fig.sketchb} in the caption of Fig.~\ref{fig:setup} and the Methods section. \\


Since inserting a beam block changes which detector can click, this interferometric device enables the following one-way communication protocol. Let Alice only have access to the detectors $D_0$ and $D_1$, and Bob only have access to the right arms of the inner MZIs. Bob can communicate to Alice remotely using the interferometer if they agree beforehand on a common timing reference to label successive time bins. To account for the high probability of photon loss and the device imperfections, Alice and Bob pre-arrange time bin durations that will contain many redundant photon attempts per bin. For each agreed-upon time bin, Bob then sets the configuration for either bit 0 or bit 1 by removing or inserting his shutter, respectively. When photons are successfully detected by one of her detectors, Alice records both which detector clicked and the time bin of the click. The redundant attempts per bin permit Alice to successfully detect at least one trial for each bin, and thus receive Bob's binary message. In principle, the communication fidelity can be made arbitrarily high for a given photon source by increasing the bin duration and thus the encoding redundancy.

It is instructive to analyze this situation by examining the forward- and backward-evolving wave functions shown in Fig.~\ref{fig.sketcha}. For a successfully transmitted 0, i.e. $D_0$ clicks in Fig.~\ref{fig.sketcha}(a), there is no overlap of the forward (red solid line) and backward (green solid line) wave functions in either the transmission channel or at Bob's site. Similarly, for a successful 1, i.e. $D_1$ clicks in Fig.~\ref{fig.sketcha}(b), there is still no overlap of the forward and backward wave functions in the transmission channel. Thus, we expect no measurable trace to be seen by weakly coupled probes placed in the transmission channel, at least to first-order in the small interaction strength \cite{Reznik2020}. 

\begin{figure}
    \centering
    \includegraphics[width = 0.8\columnwidth]{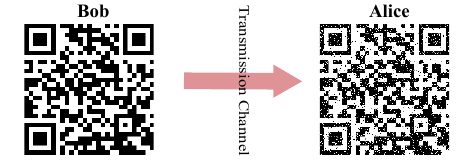}
\caption{\textbf{Demonstrating counterfactual communication.} We encoded the Chinese name of the lab 
as a QR code, raster scanned the pixels into a binary string, transmitted those bits in a counterfactual manner with the device in Fig.~\ref{fig:setup}, then converted the received bits back into a QR code image. The transmitted image is degraded from transmission imperfections, but is still readable with a common QR code reader.
}
\label{fig:qrcode}
\end{figure}

\subsection*{Results}
\vspace{-1em}\noindent \textbf{Communication demonstration.}
To test the reliability of the communication protocol we performed runs with time bins of 1000 seconds each for bit values of 1 and 0. We found error rates of 2.3\% for 1 and 10\% for 0 for the individually transmitted bits. As expected, these rates are low enough that redundant time-bin encoding could be used to eliminate logical bit transmission errors.

As a proof of principle, we then demonstrated the counterfactual transmission of a quick response (QR) code representing the Chinese name of our lab, shown in Fig.~\ref{fig:qrcode}. 
We used no redundant encoding for this demonstration, to better show the message degradation from the bare transmission error rates, and instead repeated trials until the first successful detection per bit. The QR code is a bitmap of $145\times 145$ binary pixels, so we used a linear raster scan to convert it to a string of 21025 logical bits, according to the rules:
black $\leftrightarrow$ 1 and white $\leftrightarrow$ 0. Bob then used the setup in Fig.~\ref{fig:setup} to transmit the logic bits one by one to Alice by controlling the presence or absence of the block $\mathcal{O}$. Alice obtained the bit sequence according to the clicks of her two detectors $D_0$ and $D_1$ (in coincidence with detection of the idle photons) and translated it back into the corresponding QR image shown in Fig.~\ref{fig:qrcode}, thus verifying that she successfully received information that was sent by Bob. You are welcome to scan the transmitted QR code using your own cellphone to reveal the name of our lab.

\begin{figure}
    \centering
    \includegraphics[width = 1.00\columnwidth]{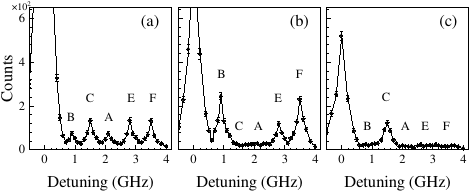}
\caption{\textbf{Frequency spectra of detected single photons.} We label each detuning peak by the letter of its corresponding EOM in Fig.~\ref{fig.sketchb}. (a) Calibration run analyzing the signal at $D_0$ in the unblocked interferometer of Fig.\,\ref{fig.sketchb}(a) with
the interferometers tuned to constructive interference towards the mirror of the MMI and constructive interference towards $D_0$, showing all peaks. (b) Sending bit 1. Signal at $D_1$ when the inner interferometer is blocked as in Fig.\,\ref{fig.sketchb}(b), with peaks for $A$ and $C$ absent and the peak heights for $B$ and $F$ doubled from the calibration height due to the double-traversal of those regions.
(c) Sending bit 0. Signal at $D_0$ when the inner interferometer is unblocked as in Fig.\,\ref{fig.sketchb}(a), with peaks at $A$, $B$, $E$ and $F$ absent.
All absent peaks are indistinguishable from the noise floor. As anticipated, these absences correspond to the regions of Fig.~\ref{fig.sketchb} with no overlap of forward- and backward-evolving wavefunctions (gray line). Notably, during successful bit transmission in (b) and (c) no peak at $A$ is ever observed, which verifies that there is no observable trace that the collected photons ever occupied the transmission channel. The errors are estimated using numerical Monte Carlo simulations considering the counting noise.
}
\label{fig.label}
\end{figure}

\noindent \textbf{Counterfactuality verification.}
We then explicitly verified that this bit transmission was indeed counterfactual according to the trace criterion, as claimed. We labeled the possible photon paths with distinct frequency shifts, then selected narrow frequency bands at the detector to resolve the spectral profiles of the collected photons.
We first prepared heralded single photons with linewidths as narrow as 300\,MHz using frequency filters as shown in Fig.\,\ref{fig:setup} (for more detail, see Methods).
We then identified possible photon locations by inserting EOMs with distinct frequency shifts in all arms of the interferometer, creating small-amplitude ($\alpha = 0.146$) sidebands, see Figs.\,\ref{fig.sketchb}(a) and \ref{fig.sketchb}(b).

Given this extra frequency information, we performed three important checks on the collected photons. First, we showed that the EOMs did correctly label the paths the photons had taken. We tuned the inner MZI to produce constructive interference towards the mirror of the MMI (not destructive as in Fig.\,\ref{fig.sketchb}(a)). We also tuned the outer MMI to produce constructive interference towards $D_0$. As seen in Fig.~\ref{fig.label}(a), we observed the expected sideband peaks corresponding to all five EOMs at $D_0$. Second, we analyzed the case of transmitting bit 1, when Bob blocks the $A$ channel of the MZI. There should be no peak at the frequency modulated by EOM-A.
The photons reaching $D_1$ should not pass through EOM-C, pass once through EOM-E and twice through EOM-B and EOM-F. These expectations correspond correctly to the peaks observed in Fig.\,\ref{fig.label}(b). Finally, we analyze the case of sending bit 0 by removing the block in arm $A$ of the MZI. The photons entering the MZI (tuned to destructive interference towards EOM-F) leave the MMI. The photons reaching $D_0$ should pass only through EOM-C, which is confirmed in Fig.\,\ref{fig.label}(c).\\

\subsection*{Discussion}
\vspace{-1em}\noindent The trace identifying where a pre- and postselected particle has gone is a subtle issue. It cannot be completely understood using Wheeler's criterion \cite{Wheeler}, in which the particle's past is determined by the trajectory of a localized wave packet (LWP). Wheeler infers the past trajectory from the forward evolving quantum wave, which can be a superposition of several LWPs, by identifying the LWP that starts at the source and ends at the detector. In most cases, Wheeler's criterion correctly tells us where a weak trace is present, but there are cases when it fails \cite{past}. 
The trace appears in the overlap of the forward and backward evolving wave functions. Note that the controversy about this approach is still not resolved. There were numerous criticisms of this approach \cite{Hasegawa,LiCom,Bart,Poto,Griff,Hash,HashComRep,Jordan,Sali,China,Nik,Sok,Berge,Bhati,Salih2022}. We, however, believe that all of them were  properly answered \cite{RepLiCom,morepast,BartCom,PotoCom,GrifRep,HashCom,JordanCom,SaliCom,ChinaCom,NikRep,SokCom,BergeCom,Wander2021,myphotonsneutrons,BhatiCom}.

We should mention that there are also publications in which other meanings of `counterfactual' are adopted  \cite{Vaidman2019}. In the context of quantum computation \cite{Jozsa}, in \cite{kong2015experimental} the computation was considered to be counterfactual when the photon that makes the computation was not absorbed during the process. In \cite{Gisin} the counterfactuality of communication corresponded to the lack of the particle in some part of the transmission channel, even if the empty parts were different for transmitting different logical bits. In \cite{arvid2017,NPJ2019} the communication protocol considered counterfactual if the particles moved in the opposite direction to the information transfer. Note that the title of \cite{NPJ2019}: `Trace-free counterfactual communication with a nanophotonic processor' is somewhat misleading, since calculations show that the trace left in the transmission channel in this setup is larger than the trace in other experiments named counterfactual and it was not measured in this experiment. `Trace-free' in \cite{NPJ2019} just meant high fidelity communication channel in which photons left very small trace which was not compared  with the trace of a single localized photon.

The  counterfactuality of the protocols we consider was also analyzed in the framework of consistent histories approach \cite{Griff,laws,SalihCH,hance2021quantum}. In particular, it was shown that there is a family of consistent histories for which the protocol with AV modification is not counterfactual \cite{laws,hance2021quantum} since one of the histories is a trajectory passing through the transmission channel.  We  do not see motivation for consistent histories   approach, and  view this argument as just one more reason against it. How helpful is it to consider this hypothetical trajectory of consistent histories when standard quantum mechanics calculations, or, a quicker TSVF consideration, show that in the laboratory  there is no trace along this trajectory? Anyway, consistent histories analysis is beyond the scope of our paper.



The weakness of our demonstration is that it is counterfactual only for detected photons and there were many other photons passing through the transmission channel: the photons of the stabilization procedure, the photons that went into the interferometer without coincidence detection of an idle photon, and most importantly the photons of the main protocol that were lost in the attenuator or that escaped the unmonitored port of the interferometer. In some cases (e.g. for quantum key distribution \cite{Noh}) there is a justification for disregarding the lost photons. In the current protocol, however, the photons which did not reach Alice's detectors still can be used (by Eve) for getting information about the transmitted message, so they cannot be disregarded. In the modification of the full counterfactual communication protocol \cite{SalihCFC,expPNAS}, we would also lose some photons (although their number can be made arbitrarily small). The lost photons leave a strong trace, but provide information only about a small fraction of transmitted bits.

The main achievement of our work is showing that the photons of successful events in a properly working counterfactual communication device, that is, properly tuned MMI with inserted MZI, Fig.\,\ref{fig.sketchb}(a) and (b), do not leave traces in the communication channel. If we are `lucky' and the first sent photon is detected by Alice in our experiment, then there will be no first order trace in the channel, while in all previous setups at least one of the logical bits cannot be transmitted without leaving such trace. We verified that the AV `patch' 
 \cite{AV19} indeed removes the traces in the communication channel. Thus, adding this patch to other counterfactual communication protocols \cite{SalihCFC,Kwiat}, including the direct transfer of quantum states \cite{SalihQCF,ZubairyQCF}, will make them truly counterfactual according to the trace definition of the presence of particles \cite{past}. This criterion is much stronger than Wheeler's classical criterion for counterfactuality \cite{Wheeler},`the particle was not in the transmission channel because if it was there, it could not be detected by Alice's detector', which is not justified when quantum particles are considered.
The modification we demonstrate here does not require new technology, so truly counterfactual communication, although challenging, is feasible with current technology.

In fact, there have already been some proposals for the implementation of the AV modification that we demonstrated here, notably, including  the ultimate task of communication of quantum states \cite{Salih2022}. Moreover, counterfactual communication task that fulfills the weak trace criterion of conterfactuality has been claimed to already have been implemented in an alternative scheme using photon polarization \cite{laws}. Although we agree that the photons in this experiment detected by Alice in her two detectors left no trace in the transmission channel, we disagree that this protocol can be called counterfactual communication of both bit values. Detection by detector $D_1$ indeed tells us in a counterfactual way that the bit is 1. However, detection of  detector $D_0$ tells us only that the bit was not tested.

\subsection*{Methods}
\vspace{-1em}\noindent\textbf{Technical improvements.} Our implementation, sketched in Fig.~3, makes a few additional modifications to the conceptual design in order to increase interference visibility and thus bit transmission fidelity. Instead of polarization-insensitive beam-splitters we used polarized light and HWPs combined with PBSs. We also used an additional laser, as well as photodetectors and piezoceramic translation stages to stabilize the phases of the MZI and the MMI, reaching final measured visibilities of 98\% and 97\%, respectively. \\

\noindent\textbf{Measurement of the trace.}
When a photon with a frequency wave packet $\Psi(\omega)$ that is well-localised around $\omega_0$ passes through a particular EOM (with modulation frequency $ \Omega_i$), its wave packet is changed to be the superposition of the original wave packet and wave packets with frequency-shifted sidebands,  $\Psi(\omega)\rightarrow\Psi(\omega)+\alpha [ \Psi(\omega +\Omega_i) + \Psi(\omega -\Omega_i)]$. The small amplitude $\alpha$ characterises the modulation strength of the EOM. Spectral analysis of the detected photons then tells us where they have been. We equip a temperature-sensitive etalon with FSR 8\,GHz and linewidth of 100\,MHz before detector so only the photons with the resonant frequency of etalon can be collected and we can obtain the spectral by changing the resonant frequency via temperature regulation. Photons that are detected in a sideband $\omega_0 +\Omega_i$ must have passed through the arm with the EOM with modulation frequency $\Omega_i$. From the presence or absence of particular sidebands we can thus infer where all photons went in a particular run. \\

\noindent\textbf{Heralded single photons with narrow linewidth.} To measure the trace of photons, a narrow-band source is necessary. We use two etalons to filter heralded single photons generated in spontaneous parametric down conversion process. Each etalon can select a series of peaks with corresponding line width and frequency interval (free spectral range) in the spectrum. When two etalons are used at the same time, only the overlapped peaks remain. We choose etalon-1 with free spectral rang (FSR) of 105\,GHz and line width of 1.4\,GHz, and etalon-2 with FSR 22\,GHz and line width 315\,MHz to pick out only one peak at central frquency ensuring a heralded single photon source with linwidth of 300\,MHz. \\

\section*{Data availability}
\vspace{-1em}
\noindent The datasets generated and analyzed during the current study are available from the corresponding author if you ask nicely.\\

\section*{Acknowledgements}
\vspace{-1em}
\noindent This work was supported by
the Innovation Program for Quantum Science and Technology (No. 2021ZD0301200)
, the National Natural Science Foundation of China (Nos.\,12022401, 62075207
, 11821404
, 11804330
, 12204142
, 12204468
),
the Fundamental Research Funds for the Central Universities (No.\,WK2470000026, 
WK2470000030
), %
the CAS Youth Innovation Promotion Association (No.\,2020447)
, Anhui Initiative in Quantum Information Technologies (Grant No.\,AHY020100
), %
the Anhui Provincial Natural Science Foundation (No.\,2208085QA13)
, the Key Program of the Education Department of Anhui Province (No.\,KJ2021A0917) 
and the U.S.-Israel Binational Science Foundation (Grant No. 735/18). JD was partially supported by National Science Foundation (NSF) (Grant No. 1915015) and Army Research Office (ARO) (Grant No. W911NF-18-1-0178).

\section*{Author contributions}
\vspace{-1em}
\noindent W.-W. P. and X. L. contributed equally to this work. W.-W. P., X. L., L.V. and X.-Y. X. designed the experiment. W.-W. P. and X. L. performed the experiment assisted by Z.-D. C., J. W., Z.-D. L. and Z.-Q. Z.. L.V. provided theoretical support. C.-F. L. and G.-C. G. supervised the project. W.-W. P., X.-Y. X., Q.-Q. W. and L.V. wrote the paper. G. C. and J. D. helped to improve the paper. All authors discussed the results and read the paper.\\

\section*{Competing interests}
\vspace{-1em}
\noindent
The Authors declare no Competing Financial or Non-Financial Interests.

\bibliographystyle{naturemag.bst}
\bibliography{references}
\vspace{16pt}

\end{document}